\documentclass[twocolumn, tighten]{aastex63}

\usepackage[single=true]{acro}
\usepackage{ulem}
\usepackage{amsmath}	

\DeclareAcronym{agn}{
  short = AGN,
  long  = active galactic nucleus,
  short-plural = s,
  long-plural-form = active galactic nuclei,
  class = astro,
  first-style = default
}
\DeclareAcronym{ebl}{
  short = EBL ,
  long  = extragalactic background light ,
  class = astro,
  first-style = default
}
\DeclareAcronym{smbh}{
  short = SMBH ,
  long  = supermassive black hole ,
  class = astro ,
  first-style = default
}
\DeclareAcronym{dsa}{
  short = DSA ,
  long  = diffusive shock acceleration ,
  class = astro ,
  first-style = default
}
\DeclareAcronym{sto}{
  short = StA ,
  long  = stochastic acceleration ,
  class = astro ,
  first-style = default
}
\DeclareAcronym{ic}{
  short = IC ,
  long  = inverse Compton ,
  class = astro ,
  first-style = default
}
\DeclareAcronym{kn}{
  short = KN ,
  long  = Klein--Nishina ,
  class = astro ,
  first-style = default
}
\DeclareAcronym{ec}{
  short = EC ,
  long  = elastic Coulomb ,
  class = astro ,
  first-style = default
}
\DeclareAcronym{pp}{
  short = \ensuremath{pp} ,
  long  = hadronuclear ,
  class = astro ,
  first-style = default
}
\DeclareAcronym{pg}{
  short = \ensuremath{p\gamma} ,
  long  = photomeson ,
  class = astro ,
  first-style = default
}
\DeclareAcronym{sed}{
  short = SED ,
  long  = spectral energy distribution ,
  class = astro ,
  first-style = default
}

\DeclareAcronym{alma}{
  short = ALMA ,
  long  = Atacama Large Millimeter/submillimeter Array ,
  class = instruments ,
  first-style = default
}

\shorttitle{On the Origin of High Energy Neutrinos from NGC~1068}

\begin{document}

\title{On the Origin of High Energy Neutrinos from NGC~1068: The Role of Non-Thermal Coronal Activity}

\author[0000-0002-7272-1136]{Yoshiyuki Inoue}
\email{yoshiyuki.inoue@riken.jp}
\affiliation{Interdisciplinary Theoretical \& Mathematical Science Program (iTHEMS), RIKEN, 2-1 Hirosawa, Saitama 351-0198, Japan}
\affiliation{Kavli Institute for the Physics and Mathematics of the Universe (WPI), UTIAS, The University of Tokyo, Kashiwa, Chiba 277-8583, Japan}

\author[0000-0002-7576-7869]{Dmitry Khangulyan}%
\email{d.khangulyan@rikkyo.ac.jp}
\affiliation{Department of Physics, Rikkyo University, Nishi-Ikebukuro 3-34-1, Toshima-ku, Tokyo 171-8501, Japan}%

\author[0000-0003-4384-9568]{Akihiro Doi}
\email{akihiro.doi@vsop.isas.jaxa.jp}
\affiliation{Institute of Space and Astronautical Science JAXA, 3-1-1 Yoshinodai, Chuo-ku, Sagamihara, Kanagawa 252-5210, Japan}%
\affiliation{Department of Space and Astronautical Science, The Graduate University for Advanced Studies (SOKENDAI),3-1-1 Yoshinodai, Chuou-ku, Sagamihara, Kanagawa 252-5210, Japan}%

\date{\today}

\begin{abstract}
NGC~1068, a nearby type-2 Seyfert galaxy, is reported as the hottest neutrino spot in the 10-year survey data of IceCube. Although there are several different possibilities for the generation of high-energy neutrinos in astrophysical sources, feasible scenarios allowing such emission in NGC~1068 have not yet been firmly defined. We show that the flux level of GeV and neutrino emission observed from NGC~1068 implies that the neutrino emission can be produced only in the vicinity of the supermassive black hole in the center of the galaxy. The coronal parameters, such as magnetic field strength and corona size, making this emission possible are consistent with the spectral excess registered in the millimeter range. The suggested model and relevant physical parameters are similar to those revealed for several nearby Seyferts. Due to the internal gamma-ray attenuation, the suggested scenario cannot be verified by observations of NGC~1068 in the GeV and TeV gamma-ray energy bands. However, the optical depth is expected to become negligible for MeV gamma rays, thus future observations in this band will be able to prove our model. 
\end{abstract}


\section{Introduction}
IceCube registered first astrophysical neutrinos with energies of TeV--PeV several years ago \citep{IceCube2013_PRL}. However, their origin remains uncertain. Recently, a nearby Seyfert galaxy NGC~1068, which appeared as the hottest spot in all-sky 10-year survey data of IceCube, was reported to be a neutrino source with 2.9-$\sigma$ confidence level \citep{IceCube2019_NGC1068}. Thus, studying possible neutrino production mechanisms in NGC~1068 is a timely task that may provide a key clue for unveiling the origin of the cosmic diffuse neutrino background flux. 

Production of very-high-energy (VHE) neutrinos is accompanied by emission of gamma-rays and the luminosity of that component exceeds the neutrino one. NGC~1068 is known as a gamma-ray emitter \citep{Lenain2010,3FHL,4FGL}.  However, the reported neutrino flux is higher than the GeV gamma-ray flux \citep{IceCube2019_NGC1068}. Thus, it requires a significant attenuation of GeV gamma-rays. Unless one adopts a very exotic spectrum of emitting particles, this implies a presence of enough dense X-ray target photons, \(\epsilon_X\sim1\rm\,keV\). The gamma-ray optical depth $\tau$ depends on the X-ray luminosity of this component $L_X$ and the size of the production regions $R$:
\begin{equation}
    \tau\approx \frac{\sigma_{\gamma\gamma}}{4\pi c}\epsilon_X^{-1}L_XR^{-1}\simeq10^5\left(\frac{\epsilon_{X}}{1~{\rm keV}}\right)^{-1}\frac{L_{X}}{L_{\rm Edd}}\frac{R_{\rm s}}{R},
\end{equation}
where $L_{\rm Edd}$ is the Eddington luminosity and $R_s$ is the Schwarzschild radius. Although this estimate is not sensitive to the mass of the source, it suggests that such a dense X-ray target can exist only in the vicinity of  compact objects. 
    
A large number, \({\cal N}\sim10^{3}\), of stellar-mass black hole (or neutron star) systems, i.e., X-ray binaries, can produce the neutrinos with the flux level required by the IceCube detection without violating the above criteria. Their typical luminosity distribution in a galaxy is represented  by a power-law as $dN/dL_X\propto L_X^{-1.6}$ \citep{Swartz2011}. This implies that the dominant contributor is the higher luminosity systems.  However, NGC~1068 hosts only three X-ray binaries with $L_X\ge10^{39}\,{\rm erg\,s^{-1}}$ \citep{Swartz2011}, thus their number is not sufficient at least by several orders of magnitude. The only remaining candidate is the \acp{smbh} at the center of the galaxy. High accretion rate, implied by the intrinsic X-ray luminosity of $L_{\rm X}\sim10^{-2}L_{\rm Edd}$ \citep{Bauer2015,Marinucci2016}, does not allow an effective particle acceleration in the central black hole magnetosphere. Thus, Emission of very-high-energy neutrinos from the vicinity of \acp{smbh} indicates the operation of efficient non-thermal particle acceleration in the \acp{agn} coronae.

NGC~1068 is a type-2 Seyfert galaxy, which is a class of \acp{agn}, emitting intense electromagnetic radiation in a broad range of frequencies. {The intrinsic X-ray emission is generated through Comptonization of accretion disk photons in hot plasma above the disk, namely in corona \citep[e.g.,][]{Katz1976,1977A&A....59..111B, Pozdniakov1977,Galeev1979,Takahara1979}. Typical size of the AGN corona is about $\gtrsim10R_s$. If high-energy particles are accelerated in the corona, the nucleus of NGC 1068 is a plausible candidate for the observed neutrinos, which is consistent with a number of previous studies \citep[see e.g.,][]{Begelman1990,Stecker1992, Kalashev2015, Inoue2019,Murase2019}.  There has been no clear observational evidence for the non-thermal coronal activity  \citep{Lin1993,Madejski1995}. However, as several corona parameters, e.g.,  magnetic field strength and corona size, remain highly uncertain, one cannot rule out the presence of high-energy particles in \ac{agn} corona. These parameters have a critical impact on the expected non-thermal flux level.

The coronal synchrotron emission is a key for dissolving this problem as it reflects the non-thermal coronal activity directly and allows determining corona magnetic properties \citep[e.g.,][]{DiMatteo1997, Inoue2014}. By combining the radio and X-ray spectra, one can also determine the size of coronae. A characteristic feature of the coronal synchrotron emission is a spectral excess in the millimeter-band, so-called mm-excess. However, previous observations had presented inconclusive evidence of such excess in the radio spectra of several Seyfert galaxies. A key observational challenge in registering the excess is the contamination by extended galactic emission and a paucity of multi-band data \citep{Antonucci1988, Barvainis1996, Doi2016, Behar2018}.

Recently, \citet{Inoue2018} reported the detection of non-thermal coronal radio synchrotron emission from two nearby Seyferts, IC~4329A and NGC~985, utilizing the \ac{alma}, which enabled multi-band observations with high enough angular resolution to exclude the galactic contamination. These observations provided the first determination of the key physical parameters of the corona: magnetic field strength and its size. Given the coronal parameters constrained by X-ray and radio observations, the non-thermal acceleration process there should be capable of boosting particle energy to the very-high-energy band resulting in the generation of high-energy gamma-ray and neutrino emission \citep{Inoue2019}.

  Very Long Baseline Array (VLBA) radio observations detected a homogeneous emission at the cm-bands from the central parsec of NGC~1068. The reported flux is attributed to the free-free emission  \citep{Gallimore2004}. This spectrum is not consistent with the fluxes observed at higher frequencies, mm-bands, which shows a spectral excess  (see the Sec.``Observational Properties''). Because of the spectral shape the excess component should be produced by non-thermal particles localized in a much more compact region, with the size similar to the one of the corona.

In this {\it Letter}, we study the constraints on the non-thermal particles in the corona of NGC~1068 imposed by the observations in the radio and gamma-ray bands. We check if the reported flux of VHE neutrino is consistent with the revealed properties of the corona non-thermal particles. We also discuss the reason why NGC~1068 appears as the hottest spot among other Seyfert galaxies based on the corona scenario. 

\section{Observational Properties}
NGC~1068 is one of the nearest and the best-studied Seyfert~2 galaxies in broadband \citep[see, e.g.,][for details]{Pasetto2019}, where the central engine is supposed to be blocked by the dusty torus. It locates at a distance of $\sim14$~Mpc \citep[$1^{\prime\prime} \sim 70$~pc,][]{Tully1988}. 

The mass of the central black hole is still uncertain. It is estimated as $\sim1\times10^7M_\odot$ from the measurement of the rotational motion of a water maser disk \citep{Greenhill1996,Hure2002,Lodato2003}. However, the rotation curve is non-Keplerian \citep{Lodato2003}. The SMBH mass is also estimated as $\sim7\times10^7M_\odot$ and $\sim1\times10^8M_\odot$ from the polarized broad Balmer emission line and the neutral FeK$\alpha$ line, respectively \citep{Minezaki2015}. In this study, we adopt $5\times10^7M_\odot$ as the mass of the central SMBH of NGC~1068.

In centimeter radio observations, the jets are prominent and extend for several kpcs in both directions. In the central $\sim1''$ region, the downstream jet emission dominates in the centimeter regime. The jet changes its direction at $\sim0.2''$ away from the nuclear region. This change is presumed to be the result of an interaction with a molecular cloud \citep{Gallimore1996, Gallimore2004, Cotton2008}. At long wavelengths, the VLBA 1.4~GHz, 5~GHz, and 8.4~GHz (``cm-bands'') observations reported the flux of $5.9\pm0.5$~mJy with a spectral index of $-0.17$ above 5~GHz and nondetection at 1.4~GHz, indicating strong attenuation below 5~GHz \citep{Gallimore2004}. At 5~GHz, the brightness temperature is $\sim2.5\times10^6$~K, which is too low for synchrotron self-absorption flux unless the magnetic fields are order of $10^9$~G \citep{Gallimore1996, Gallimore2004}. As discussed in \citet{Gallimore2004}, the free-free emission is the most likely origin of the 5~GHz and 8.4~GHz flux. The emission region size is $\sim0.8$~pc with a temperature of $\sim10^6$~K and the electron density of $\sim8\times10^5\ {\rm cm}^{-3}$, which is irrelevant to the central corona and much more extended. At short wavelengths (``mm-bands''), the central compact region starts to dominate the entire emission \citep{Cotton2008, Imanishi2018}. Together with multi-frequency observations, a possible spectral excess was reported in the nucleus component \citep{Krips2006, Pasetto2019}. Recently, high angular resolution observations with \ac{alma} detected  $6.6\pm0.3$~mJy and $13.8\pm1.0$~mJy flux from the core at 256~GHz and 694~GHz, respectively \citep{GarciaBurillo2016, Impellizzeri2019}. As the expected free-free component at the mm band based on the VLBA observations is only at $\sim4$~mJy, a new spectral component is emerging in the mm band. 

In X-rays, NGC~1068 was first studied by {\it Ginga} \citep{Koyama1989}, where intense iron line was reported together with an estimate for the intrinsic X-ray luminosity of $10^{43-44}\ {\rm erg\ s^{-1}}$. Later, it is reported that the observed X-ray emission is due to the reflected component \citep[e.g.,][]{Ueno1994}. Utilizing {\it NuSTAR} and other existing X-ray observatories, the reflected emission is revealed to be originated in multiple reflection components \citep{Bauer2015, Marinucci2016}. The intrinsic 2-10~keV luminosity is estimated as $L_{\rm X}=7^{+7}_{-4}\times10^{43}\ {\rm erg\ s^{-1}}$ \citep{Marinucci2016}, while according to \citet{Bauer2015} it is $2.2\times10^{43}\ {\rm erg\ s^{-1}}$. In this {\it Letter}, we take the value of $L_{\rm X}=7\times10^{43}\ {\rm erg\ s^{-1}}$ as the fiducial value.

\begin{figure}[t]
 \begin{center}
  \includegraphics[width=\linewidth]{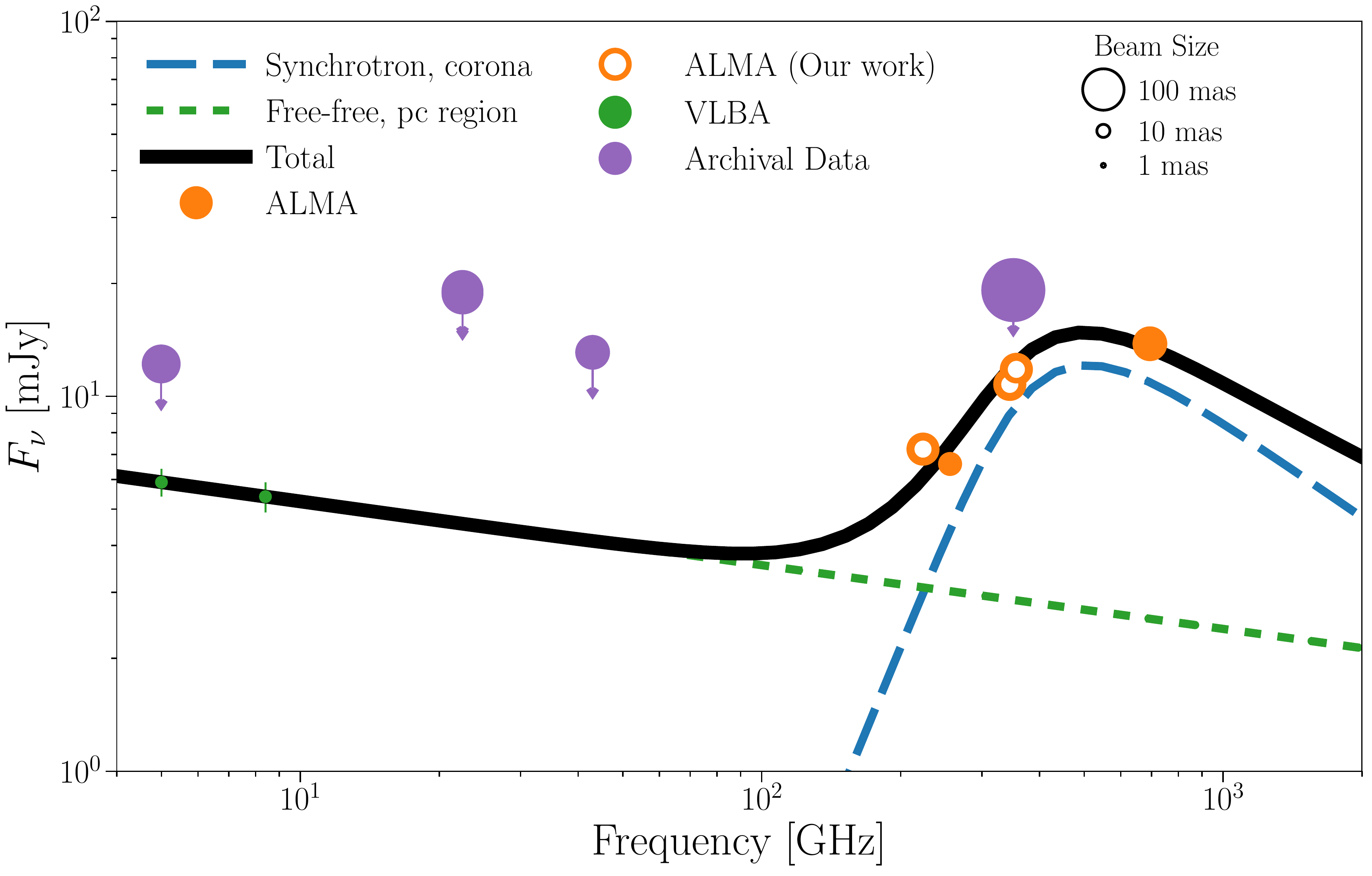}
\caption{The cm-mm spectrum of NGC~1068. {The data points from VLBA \citep{Gallimore2004} and ALMA \citep{GarciaBurillo2016,Impellizzeri2019,GarciaBurillo2019} are shown in green and orange, respectively. The open points represent the newly analyzed ALMA data. The size of circles corresponds to the beam sizes as indicated in the figure.} {We also show the archival mm-cm data having large beam sizes as upper limits in purple \citep{Gallimore1996,Cotton2008,Pasetto2019}.} The error bars correspond to 1-$\sigma$ uncertainties{, although those can not be clearly seen because of their small errors}. The blue-dashed and green-dotted lines show the coronal synchrotron and {pc-scale} free-free component, respectively. The black solid line shows the sum of these two components.}\label{fig:ALMA}
 \end{center}
\end{figure}

\section{Coronal Synchrotron Emission}
Fig. \ref{fig:ALMA} shows the cm-mm spectrum of NGC~1068 based on measurements reported by \citet{Gallimore2004,GarciaBurillo2016,Impellizzeri2019}{, where the beam size is $\sim2$, $\sim50$, $\sim20$~mas, respectively. In addition, we obtain the continuum fluxes at 224, 345, and 356~GHz with beam sizes of 30~mas by analyzing the latest ALMA band 6 and 7 data \citep[2016.1.00232.S,][]{GarciaBurillo2019}}.   {We also show the archival core region data sets having beam sizes $>50$~mas as upper limits \citep{Gallimore1996,Cotton2008,Pasetto2019}}. 

NGC~1068 shows a mm-excess similar to those observed in previous objects \citep{Inoue2018}. {However}, we cannot claim a firm detection of this component in NGC~1068, {because of a paucity of flux measurements, mixture of beam sizes, and the complex source structure}. {In this letter, motivated by the possible neutrino detection \citep{IceCube2019_NGC1068} and detection of coronal synchrotron emission in other Seyferts \citep{Inoue2018}, we consider specifically the possibility of coronal synchrotron emission to explain this mm-excess. The coronal synchrotron self-absorption, breaking the spectrum  between \(300\) and \(600\)~GHz, would imply a firm relation between the corona magnetic field and size, putting constraints on the acceleration process at work in the corona.}

The excess of NGC~1068 {can be reproduced} with the coronal synchrotron emission model with parameters of the corona size $R_c=10~R_s$, the magnetic field strength $B=100$~G, and the spectral index of non-thermal electrons $p=2.7$ {on top of free-free emission produced in pc-size region \citep{Gallimore2004}}. A much softer spectral index would violate either the {ALMA} measurements or the flux upper limit at 351~GHz (Fig.~\ref{fig:ALMA}). The required coronal size is consistent with optical--X-ray spectral fitting studies \citep{Jin2012} and micorolensing observation \citep{Morgan2012} in other Seyferts.  The previously reported coronal synchrotron objects IC~4329A and NGC~985, whose black hole masses are $\sim10^8M_\odot$, have $R_c\sim40~R_s$ and $B\sim10$~G \citep{Inoue2018}, which are similar to what we required for the excess in NGC~1068 for the coronal synchrotron emission scenario. 

Although we chose the coronal synchrotron emission model as a possible explanation for the mm-excess, current data sets do not allow us to give a clear evidence. Future {ALMA observations will be able to elucidate the origin of the excess. High angular resolution observations will be able to precisely understand the contribution of the free-free component.} Multi-frequency {and variability measurements} in the mm-band will be {also} able to {determine the spectral shape and see the coronal activity}.

The coronal synchrotron emission is mostly determined by the following parameters: $R_c$, $B$, $p$, and the energy fraction of non-thermal electrons ($f_{\rm nth}$). We fix the energy fraction of non-thermal electrons as $f_{\rm nth}=0.03$, which is required to explain the cosmic MeV gamma-ray background radiation by the Comptonization counterpart \citep{Inoue2019}. We adopt standard coronal temperature and Thomson scattering optical depth value of $100$~keV and $1.1$, respectively \citep{Inoue2019}, because both of them are not determined in \citet{Marinucci2016}.

\section{Coronal Gamma-ray and Neutrino Emission}
We investigate the properties of high energy emission from the nucleus of NGC~1068 utilizing the derived coronal parameters based on the possible mm excess (See Fig.~\ref{fig:ALMA}). Observations of the electromagnetic emission constrain the parameters of the gamma-ray and neutrino production model \citep{Inoue2019} except for the energy injection ratio between protons and electrons and the gyro factor $\eta_g$, which is the mean free path of a particle in units of the gyroradius.

Particles are expected to be accelerated by the diffusive shock acceleration. Other mechanisms such as magnetosphere, turbulence, or reconnection can not explain the required electron distribution for the coronal synchrotron emission because of high accretion rate and low magnetic field strength (See Sec.~8.3 in \citet{Inoue2019} for details). Gamma-rays are generated through the Comptonization of disk photons and/or hadronic interactions in coronae. Hadronic interactions generate neutrinos. Because of the intense X-ray and UV photon field from the corona and the accretion disk, $\gtrsim100$~MeV gamma-rays are significantly attenuated. 

Fig. \ref{fig:GN_1} shows the expected gamma-ray and neutrino signals from NGC~1068 together with the observed gamma-ray data \citep{4FGL,3FHL,MAGIC2019} and the IceCube data \citep{IceCube2019_NGC1068}. The data is taken from Fig.~7 of \citet{IceCube2019_NGC1068}, and the 1, 2, and 3 $\sigma$ regions are shown in the plot. For the comparison, we also show the expected sensitivity curves of a future MeV gamma-ray mission, {\it GRAMS} \citep{Aramaki2019} and {\it AMEGO} \citep{AMEGO}. Gamma-ray model curve is the summation of leptonic and hadronic gamma-rays after internal (X-ray corona and UV accretion disk photons)  and intergalactic (on EBL photons)  attenuation. For hadronic processes, we consider both $pp$ and $p\gamma$ interactions.

We follow the assumptions on the coronal parameters as in \citet{Inoue2019} except for the gyrofactor and parameters determined by the coronal synchrotron model explaining the mm-excess. Considering the measurement uncertainty, in the figure, we plot the model curve region in the range of $30\le\eta_g\le3\times10^4$ for each curve. The darker region corresponds to lower $\eta_g$, in which models extend to higher energies. The injection powers both in protons and electrons are set to equal as assumed in \citet{Inoue2019}. 

The gyrofactor $\eta_g=30$ was required in order to explain the measured TeV diffuse neutrinos \citep{Inoue2019}. We note that  the model curve with this $\eta_g$ is still acceptable, given the measurement uncertainty (according to the Fig.4 of \citet{IceCube2019_NGC1068}, the neutrino spectral index is constrained to a broad range above \(2\)). Further detailed neutrino spectrum will allow us to narrow down the range of $\eta_g$. If future data requires $\eta_g\gg 30$, it will suggest that Seyferts are sub-dominant contributors to the diffuse neutrino background fluxes.

\begin{figure}[t]
 \begin{center}
  \includegraphics[width=\linewidth]{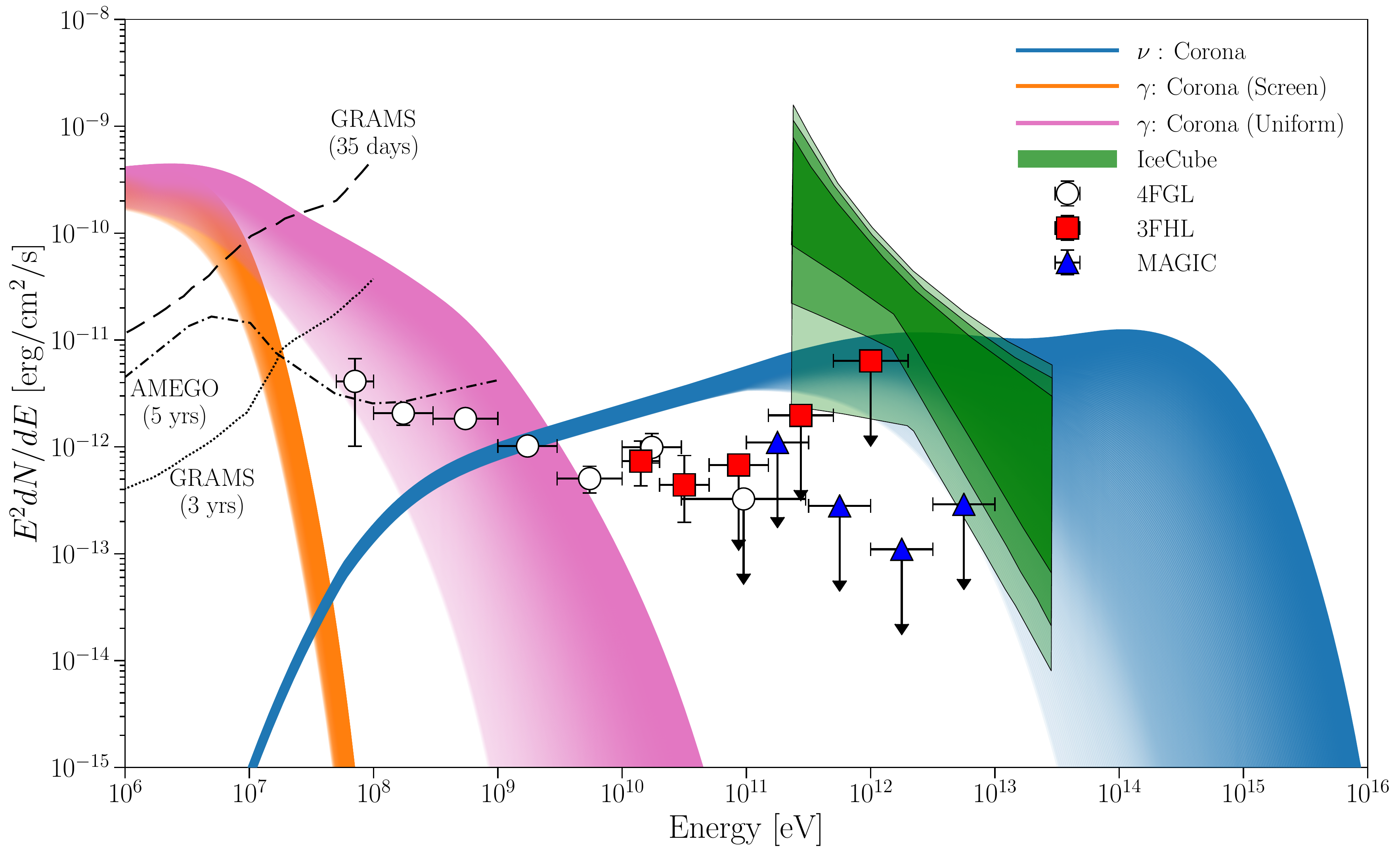}
\caption{The gamma-ray and neutrino spectrum of NGC~1068. The circle, square, and triangle data points are from \citet{4FGL}, \citet{3FHL}, and \citet{MAGIC2019}, respectively. The green shaded regions represent the 1, 2, and $3\sigma$ regions on the spectrum measured by IceCube  \citep{IceCube2019_NGC1068}. The expected gamma-ray and neutrino spectrum from the corona are shown for $30\le\eta_g\le3\times10^4$. The darker region corresponds to lower $\eta_g$. The blue region shows the expected neutrino spectrum. The orange and magenta shaded region shows the gamma-ray spectrum for the uniform case and the screened case, respectively. We also overplot the sensitivity curves of {\it GRAMS} \citep{Aramaki2019} and {\it AMEGO} \citep{AMEGO} for for comparison.}\label{fig:GN_1}
 \end{center}
\end{figure}

As NGC~1068 is an AGN, it also features a bright accretion disk and a hot corona. The disk and corona emission absorbs gamma rays via pair creation. For the internal gamma-ray attenuation, we consider two cases. One is the ``uniform'' emissivity case as assumed in \citep{Inoue2019}, while the other is the ``screened'' case. In the uniform emissivity case, gamma-rays and target photons are uniformly distributed. Gamma rays are attenuated by a factor of $3u(\tau)/\tau$, where $u(\tau)=1/2 + \exp(-\tau)/\tau - [1 - \exp(-\tau)]/\tau^2$. Here $\tau$ is the gamma-ray optical depth computed from the center of the corona. In the screened case, gamma-rays are assumed to be generated in the inner part of the corona, and the dominant attenuating photon field surrounds it. Since the disk and corona temperature depends on the disk radius \citep{Kawanaka2008}, such configuration can be realized. Then, gamma-rays are attenuated by a factor of $\exp(-\tau)$. Gamma-rays are also attenuated by the intergalactic photons during the propagation to the Earth. In this {\it Letter}, we adopt \citet{Inoue2013} for the intergalactic attenuation.

In the screened case, the model can explain the preliminary neutrino signals above several~TeV without violating the gamma-ray data. On the other hand, the uniform emissivity model violates the low-energy gamma-ray data. This implies a further detailed study of the coronal geometry is necessary. Future MeV gamma-ray missions such as {\it GRAMS} \citep{Aramaki2019} and {\it AMEGO} \citep{AMEGO} will verify our model and help us to understand the coronal geometry, which is not well understood yet. 

Due to the internal attenuation, it is not easy for the corona model to explain the entire observed gamma-ray flux data up to 20~GeV, requiring another mechanism to explain gamma-rays above 100~MeV such as star formation activity \citep{Ackermann2012_SB}, jet \citep{Lenain2010}, or disk wind \citep{Lamastra2016}. 
 
\section{Discussions and Conclusion}
The IceCube collaboration reported NGC~1068 as the hottest spot in their 10-year survey \citep{IceCube2019_NGC1068}. Surprisingly, the reported neutrino flux is higher than the GeV gamma-ray flux, which requires different origins and a significant attenuation of GeV gamma-rays from the neutrino production site. This further implies a presence of enough dense X-ray target photons in the neutrino production region in order to attenuate gamma-rays $\gtrsim100$~MeV. Such a dense X-ray target can exist only in the vicinity of compact objects. However, stellar-mass objects such as X-ray binaries can not explain the whole neutrino flux because the number of such objects in NGC 1068 is several orders of magnitude fewer than requirement. The only feasible candidate is the coronal activity of \acp{smbh} at the center of the galaxy.

NGC~1068 is one of the best-studied type-2 Seyfert galaxies. The nucleus flux in the cm band comes from the  free-free emission component \citep{Gallimore2004}. However, at higher frequencies, an excess of core flux is reported utilizing \ac{alma} \citep{GarciaBurillo2016, Impellizzeri2019}. We found that the coronal synchrotron emission model can reproduce the observed mm spectrum, which puts constraints on the acceleration process in the corona. 

Given the corona parameters revealed with \ac{alma} measurements, we studied the resulting gamma-ray and neutrino emissions from the corona of NGC~1068. Although it is difficult to explain the gamma-ray flux above 100~MeV due to significant internal attenuation effect, the coronal emission can explain the reported IceCube neutrino flux with the gyro factor in the range of $30\le \eta_g \le 3\times10^4$. Further neutrino data on NGC~1068 will narrow down the required range of $\eta_g$. It should be noted that $\eta_g\sim30$ is required for Seyferts to explain the diffuse neutrino fluxes up to 300~TeV \citep{Inoue2019}.

In order not to violate the observed gamma-ray data, the corona can not be uniform. The dominant attenuating photon field needs to surround the gamma-ray emission region. Since the disk temperature depends on the disk radius, such a configuration can be realized. Future MeV gamma-ray observations will be the critical tool to test the corona scenario.

An important question is what differs NGC~1068 from other nearby Seyfert galaxies. NGC~1068 is not the brightest X-ray Seyfert \citep{Oh2018}. Its observed hard X-ray flux is a factor of $\sim16$ fainter than the one of the brightest Seyfert, NGC~4151. NGC~1068 is a type-2 Seyfert galaxy, and obscured by the materials up to the neutral hydrogen column density of $N_H\sim10^{25} {\rm cm^{-2}}$ \citep{Bauer2015, Marinucci2016}. If we correct this attenuation effect to understand the intrinsic X-ray radiation power, NGC~1068 appears to be the intrinsically brightest Seyfert. For example,  intrinsically, it would be by a factor of $\sim3.6$ brighter than NGC~4151 in X-ray. As the dusty torus does not obscure coronal neutrino emission, which can scale with accretion power, NGC~1068 might be the brightest source of VHE  neutrinos. This could be the reason why NGC~1068 appears as the hottest spot in the IceCube map rather than other Seyfert galaxies.

\acknowledgements
The authors thank the anonymous referee for thoughtful and helpful comments. YI is supported by JSPS KAKENHI Grant Number JP16K13813, JP18H05458, JP19K14772, program of Leading Initiative for Excellent Young Researchers, MEXT, Japan, and RIKEN iTHEMS Program. DK is supported by JSPS KAKENHI Grant Numbers JP18H03722, JP24105007, and JP16H02170. {This paper makes use of the following ALMA data: ADS/JAO.ALMA\#2016.1.00232.S. ALMA is a partnership of ESO (representing its member states), NSF (USA) and NINS (Japan), together with NRC (Canada), MOST and ASIAA (Taiwan), and KASI (Republic of Korea), in cooperation with the Republic of Chile. The Joint ALMA Observatory is operated by ESO, AUI/NRAO and NAOJ.}

\bibliography{references}{}
\bibliographystyle{aasjournal}

\listofchanges

\end{document}